\documentclass[]{fairmeta}

\usepackage[T1]{fontenc}
\usepackage[utf8]{inputenc}
\usepackage{lmodern}

\makeatletter
\DeclareFontFamily{T1}{optimistic}{}
\DeclareFontShape{T1}{optimistic}{m}{n}{<-> ssub * lmr/m/n}{}
\DeclareFontShape{T1}{optimistic}{m}{it}{<-> ssub * lmr/m/it}{}
\DeclareFontShape{T1}{optimistic}{bx}{n}{<-> ssub * lmr/bx/n}{}
\DeclareFontShape{T1}{optimistic}{bx}{it}{<-> ssub * lmr/bx/it}{}
\makeatother

\usepackage{amsmath,amssymb,amsfonts,amsthm}

\usepackage{graphicx}
\usepackage{multirow}
\usepackage{adjustbox}
\usepackage{booktabs}

\usepackage{algorithmic}
\usepackage{textcomp}
\usepackage{xcolor}
\usepackage{paralist}
\usepackage{balance}
\usepackage{tablefootnote}
\usepackage{wasysym}
\usepackage{microtype}

\newcommand{\sectionname}{Section}
\newcommand{\equref}[1]{Equation (\ref{#1})}
\newcommand{\para}[1]{\noindent\textbf{#1.}}
\newcommand{\subpara}[1]{\smallskip\noindent\textit{#1.}}
\newcommand{\sln}{HRNG}

\title{Adding All Flavors: A Hybrid Random Number Generator for dApps and Web3}

\author[2]{Ranjith Chodavarapu}
\author[1\dagger]{Rabimba Karanjai}
\author[3]{Xinxin Fan}
\author[1]{Weidong Shi}
\author[2]{Lei Xu}

\affiliation[1]{University Of Houston}
\affiliation[2]{Kent State University}
\affiliation[3]{IoTex}
\contribution[\dagger]{rabimba@cs.uh.edu}

\abstract{
    Random numbers play a vital role in many decentralized applications (dApps), such as gaming and decentralized finance (DeFi) applications.
    Existing random number provision mechanisms can be roughly divided into two categories, on-chain, and off-chain. 
    On-chain approaches usually rely on the blockchain as the major input and all computations are done by blockchain nodes.
    The major risk for this type of method is that the input itself is susceptible to the adversary's influence.
    Off-chain approaches, as the name suggested, complete the generation without the involvement of blockchain nodes and share the result directly with a dApp.
    These mechanisms usually have a strong security assumption and high complexity.
    To mitigate these limitations and provide a framework that allows a dApp to balance different factors involved in random number generation, we propose a hybrid random number generation solution that leverages IoT devices equipped with trusted execution environment (TEE) as the randomness sources, and then utilizes a set of cryptographic tools to aggregate the multiple sources and obtain the final random number that can be consumed by the dApp.
    The new approach only needs one honest random source to guarantee the unbiasedness of the final random number and a user can configure the system to tolerate malicious participants who can refuse to respond to avoid unfavored results.
    We also provide a concrete construction that can further reduce the on-chain computation complexity to lower the cost of the solution in practice.
    We evaluate the computation and gas costs to demonstrate the effectiveness of the improvement.
}

\date{20 October 2024} 

\begin{document}

\maketitle

% Abstract and keywords

% Introduction

%\xinxin{This blogpost might be helpful for writing abstract and introduction \url{https://chain.link/education-hub/randomness-web3}}
\section{Introduction}
Random numbers play an important role in a wide range of applications, such as machine learning, simulation, and cryptography.
When it comes to blockchain, random numbers are critical for not only the blockchain construction itself (e.g., implementation of proof-of-stake), but also dApps deployed on top of it (e.g., gaming and NFT).
An adversary can gain extra advantages if it can manipulate or influence the random numbers involved.

A good random number, or a good random number sequence, should meet two requirements:
\begin{inparaenum}[(i)]
    \item \textit{Unpredictable}. One should not learn the value before it is released or guess future a random number from historical information; and
    \item \textit{Unbiased}. The value should follow the uniform distribution, i.e., each value is selected with the same probability;
\end{inparaenum}
These two requirements are not independent. It is easy to see that if the random numbers generated are biased, it is easier for the adversary to predict it.
When the random number is generated in a hostile environment, there are extra security related requirements.
For instance, an adversary should not be able to influence the value directly/indirectly.
In the context of blockchain and dApp, it is also desirable that the random number generation is verifiable by all participants.
In this paper, we focus on random number generation used in a decentralized environment, especially dApps.

%random numbers do not meet these requirements can give the adversary extra advantages in winning a game, minting NFTs. 
Random number generation can be classified by various criteria.
Based on the original source, random number generators can be divided into \textit{pseudo random number generator (PRNG)} and \textit{true random number generation (TRNG)}.
A TRNG utilizes unpredictable physical sources to generate random numbers, and a PRNG uses a mathematical algorithm and a seed value to generate a sequence of numbers.
While it is widely believed that TRNG offers random numbers with better quality, most random numbers used in practice are created by PRNGs, because TRNG requires hardware that is not always available.
From the perspective of dApps, the random number generation mechanisms are usually divided into two categories, on-chain RNG and off-chain RNG.
An on-chain RNG usually uses the blockchain information as a seed to produce random numbers.
This approach is convenient for the dApp to verify and consume the generated random number but the seed is suspectable to the influence of an adversary.
On the other hand, an off-chain RNG does not rely on the blockchain contents to generate random numbers but the process is usually opaque to the dApp and the dApp participants need to trust that this process is not manipulated.

%Compared with traditional applications, a dApp has extra requirements on the random number.
%For instance, the random number must be publicly verifiable, i.e., all participants need to be convinced that the random number used in the dApp is correct.
%Another requirement is that everyone should be convinced that that random number generator does not collude with the dApp who consumes the random numbers, because a colluded random number generator may discard unfavored numbers and only repeat the generation process until it gets one that the dApp prefers.

% idea of the the solution
To better serve the demands of dApps on random numbers, we propose \sln{}, a hybrid RNG solution that keeps the advantages of all four types of RNGs while minimizing related limitations.
\sln{} takes advantage of the emerging DePIN technology (decentralized physical infrastructure network~\cite{ballandies2023taxonomy}) to solve the problem of introducing physical randomness sources to the blockchain world.
The idea of DePIN is to connect a large number of IoT devices to the blockchain and allow them to trade their generated data on the blockchain~\cite{xu2021new}.
When an IoT device is equipped with a hardware-based trusted execution environment (TEE), it can utilize the hardware to work as a TRNG.%~\cite{zhao2014providing}. 
If all nodes of the blockchain trust the IoT device, the problem of obtaining a random number becomes trivial.
For instance, the dApp can request a random number from the IoT device, and the TEE hardware of the IoT device responds with a generated number and a digital signature.
As the device is trusted, one only needs to verify whether the provided number and the digital signature are consistent, and does not need to worry that the device colludes with the dApp.

However, as a hardware-aided security enhancement mechanism, TEE is not 100\% reliable.
In other words, it is possible that an adversary compromises the TEE hardware and manipulates the generation of random numbers~\cite{mutlu2019rowhammer}.
The good thing is that existing attacks against TEE are usually not scalable, i.e., it is hard to compromise a large number of IoT devices equipped with TEE simultaneously.
Therefore, \sln{} utilizes a group of TEE-enabled IoT devices and aggregates their outputs to provide random numbers to dApps to tolerate the potential risk of compromised hardware.
Specifically, \sln{} uses multiple TEE devices to create a pool of random numbers, and allows a dApp to combine multiple values from the pool using a preferred method to obtain the final random number it needs as long as it meets the security requirement.
Even if some the input random numbers are manipulated by an adversary, the combination mechanism can still guarantee that the final result is not biased.
Compared with existing random number generation mechanisms for dApps, \sln{} provides the following benefits:
\begin{inparaenum}[\bf (i)]
    \item \textit{TRNG-based}. \sln{} utilizes TRNGs as the source to build the final random numbers for dApps, and it enjoys all the advantages of TRNG.
    %\item \emph{Randomness}: The dApp cannot predict the generated number before the number is released to the public;
    \item \textit{Robustness}. \sln{} can not only tolerate IoT devices with compromised TEE, but also other malicious participants in the random number generation process under reasonable security assumptions.
    %If a subset of TEE devices are compromised, they cannot collude to make the final random number biased; 
    %\item \emph{Fairness}: The dApp cannot refuse a random number that is generated following the \sln{} protocol; and
    \item \textit{Verifiability}. Except that any third party can verify that the final random number that is consumed by the dApp is obtained following the predefined aggregation method.
\end{inparaenum}

In summary, our contributions in the paper include:
\begin{inparaenum}[(i)]
    \item We provide the details of the design of \sln{} that utilizes TRNGs as a building block to generate random numbers for dApps. The design is generic and highly customizable, i.e., one can easily instantiate the system with preferred cryptographic tools based on the demands; and 
    \item We propose a concrete construction of \sln{} that utilizes the homomorphic feature of the underlying commitment scheme to significantly reduce the on-chain computation complexity, which makes the random numbers more affordable for dApps.
\end{inparaenum}

% Background and related works

\section{Background and Related Works}\label{sec-background}
%In this section, we briefly review existing random number generation mechanisms for blockchain and dApps, and related works.
Existing random number generation mechanisms for dApps are roughly divided into two categories, off-chain random number generation and on-chain random number generation.
In this section, we briefly review these mechanisms and compare them with \sln{}.

\para{Off-chain random number generation}
Using an off-chain source for random number generation is the most convenient approach from an engineering perspective. Randao~\cite{randao} is a random number generation mechanism used in the beacon chain to provide in-protocol randomness in Ethereum 2.0~\cite{randaoethereum}. The design of Randao follows the commit-reveal approach to incrementally gather randomness from participants. Motivated by the design of RanShare~\cite{7958592}, the Near protocol described a randomness beacon scheme~\cite{nearbeacon} that inherits the randomness properties of Randshare and can tolerate up to 2/3 malicious actors before one can influence the output. Chainlink built a provably fair and verifiable random number generator based on verifiable random functions~\cite{chainlinkvrf}. Chainlink VRF relies on a decentralized oracle network and can generate one or multiple random numbers together with a cryptographic proof in a single smart contract call. The proof, which can attest to the correctness of the random number generation process, is published and verified on-chain before a dApp consumes the generated random numbers.  
The off-chain collaborative random number generation schemes like Randao often require multiple participants and communication rounds, thereby incurring significant costs and delays. The VRF-based approach, on the other hand, relies on strong protection of private keys used for computing VRFs.    

%Verifiable delay functions (VDFs)~\cite{10.1007/978-3-319-96884-1_25} and BLS threshold signatures~\cite{difinity}

\para{On-chain random number generation}
Creating random numbers on-chain is another option.
One straightforward approach for on-chain random number generation is using the block contents (e.g., block headers, block heights, timestamps, smart contract states, etc.) as the random source~\cite{8751326,10174962}. The ERC721R~\cite{erc721r,erc721rupdate} standard describes the process for producing random numbers using the aforementioned on-chain metadata. While it is easy for a smart contract to access various on-chain data, it is extremely difficult, if not impossible, to prevent on-chain random number generators from being exploited by attackers~\cite{8751326,10174962}.

\begin{table}
    \centering
    \caption{Comparison of \sln{} with existing random number generation methods.}
    \label{tbl-comparison}
    \begin{tabular}{p{0.7in}p{1in}p{1in}p{1in}}
        \toprule
        \textbf{Method}  & \textbf{True Random source}  & \textbf{Verifiability} & \textbf{Tolerance of adversaries}\\
        \midrule
        On-chain    &   \Circle    & \CIRCLE &  \LEFTcircle\\
        \hline
        Off-chain   &   \Circle  &  \LEFTcircle  &    \LEFTcircle\\
        \hline
        \sln{}  &   \CIRCLE &   \CIRCLE &  \CIRCLE \\
        \bottomrule
    \end{tabular}
    
    \smallskip
    {\CIRCLE: Yes \quad \Circle: No \quad \LEFTcircle: Partially}
\end{table}

\para{This work}
Compared with existing on-chain and off-chain random number provision methods, \sln{} provides a new way to split the on-chain and off-chain workload, and utilizes TEE technology to incorporate TRNG into the process of random number generation.
Furthermore, \sln{} also considers the associated centralization risk in the design.
\tablename~\ref{tbl-comparison} summarizes the comparison of these methods, and more detailed analyses are provided in later sections.

% \LEFTcircle   \Circle     \CIRCLE

% Overview, security assumption, problem statement

\section{System Overview, Security Assumptions and Problem Statement}\label{sec-sln}
In this section, we first present the overview of the decentralized pseudo-random number generation with TEE technology, and then explain the technique challenges. 

\begin{figure}
    \centering
    \includegraphics[width=0.5\linewidth]{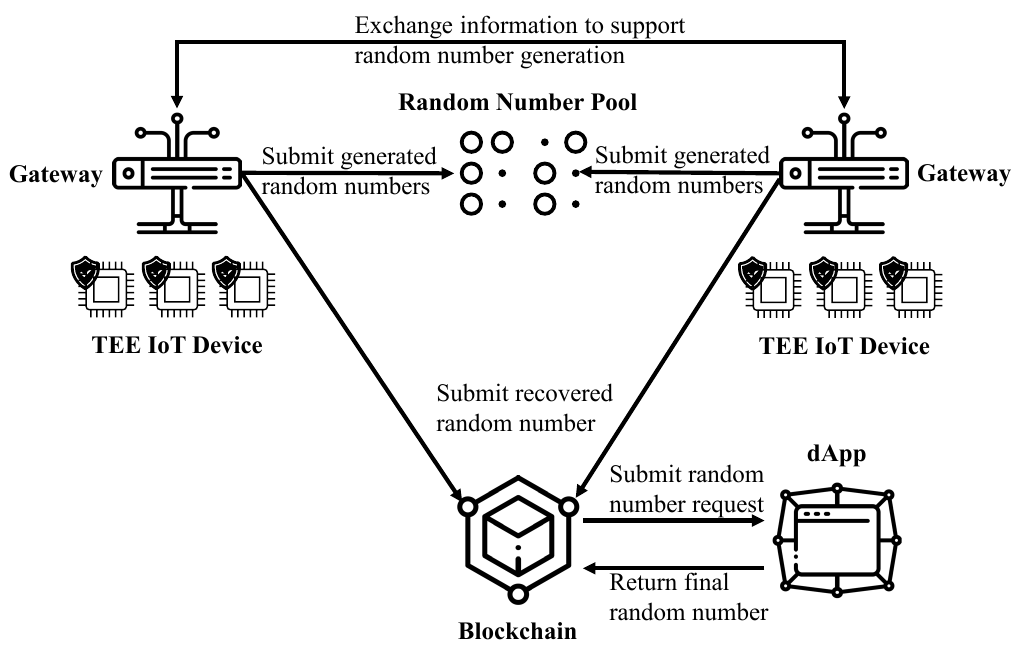}
    \caption{Overview of \sln{}. Each gateway connects with multiple IoT devices equipped with TEE to provide random sources to the system, and the final random number is generated in a verifiable manner on the blockchain. 
    Although the dApp is demonstrated as a different entity, it is actually deployed on the same blockchain and maintained by the same set of blockchain nodes.}
    \label{fig-iot-tee-prng}
\end{figure}

\para{Overview of \sln{}}
\figurename~\ref{fig-iot-tee-prng} shows the major participants involved in the random number generation process.
\begin{inparaenum}[\bf (i)]
    \item \textit{IoT device.} An IoT device utilizes equipped TEE to create random numbers to contribute to the provision of final random numbers to dApps. An IoT device only interacts with the gateway it connects with.
    \item \textit{Gateway.} A gateway works as the proxy of a group of IoT devices. The gateway has more computation/communication capacity. It can relay messages generated by connected IoT devices to other destinations and process received IoT messages before sending them out.
    \item \textit{Random number pool.} The random pool is a storage service that holds information submitted by gateways and allows the blockchain nodes/gateways to retrieve information.
    \item \textit{dApp.} A dApp is an application deployed on the blockchain, and its execution requires a random number as input. Different random numbers may lead to completely different execution results.
    %\item Random number aggregator. The random number aggregator is responsible for constructing random numbers from the random numbers pool based on the request of the dApp.
    \item \textit{Blockchain.} The blockchain serves as an immutable ledger to store the necessary information and it is also responsible for the execution of certain computations to support the operation of the whole protocol.
\end{inparaenum}

The execution of the protocol consists of four major steps:
\begin{inparaenum}[\bf (i)]
    \item \textit{Publish the random numbers pool.} During this step, each gateway collects random numbers generated by connected IoT devices, processes them, and publishes processing results to the pool;
    \item \textit{Obtain instructions from the dApp.} The dApp creates a request for a random number, which includes specifications such as the number of sources and the aggregation method;
    \item \textit{Respond to the dApp request on a random number.} The gateways obtain the request of the dApp, and work together to generate responses, which are published to the blockchain; and
    \item \textit{Construct the final random number.} The blockchain (through the smart contract deployed on the blockchain) follows the request provided by the dApp to build the final random number, which can be consumed by the dApp. 
\end{inparaenum}
Note that the execution of the protocol follows the sequence strictly, i.e., one step only starts when the previous one finishes.

\para{Security assumptions}
The major security concern is that an adversary influences the random number generation process and forces the protocol to generate a biased number based on its preference, which may cause various consequences depending on the nature of the dApp.
In the extreme case, the adversary controls all components of the system (i.e., all gateways and connected IoT devices), and there is no way to prevent it from choosing a preferred random number.
In this work, we consider a more practical scenario where the adversary can only compromise a fraction of each type of participant. Specifically,
\begin{inparaenum}[\bf (i)]
    \item \textit{Compromised IoT devices}. We assume hardware-aided TEE is breakable but hard to compromise at a large scale. Although we assume each IoT device is equipped with TEE, we consider the case that the TEE hardware can be compromised and the IoT device provides biased random numbers or refuses to. At the same time, we assume that only a subset of IoT devices in the system are compromised.
    \item \textit{Compromised gateways}. We assume a subset of gateways in the system are malicious but not all of them. A malicious gateway may refuse to respond to the system and/or deviate from the protocol to maximize the benefit of the colluded dApp.
    \item \textit{Trusted random number pool}. We assume the random number pool is a reliable storage system, i.e., it is always available and can safely store received information. 
    \item \textit{Trusted blockchain}. Following the common assumption of a blockchain system, we assume that some blockchain nodes can be malicious but the blockchain is trusted as a whole. This also applies to the dApp deployed on the blockchain.
\end{inparaenum}

For ease of description, we summarize notations that are used in the rest of the paper in \tablename~\ref{tbl-notations}.
\begin{table}
    \centering
    \caption{Summary of notations.}
    \label{tbl-notations}
    \begin{tabular}{p{0.7in}p{4in}}
        \toprule
        \textbf{Notation}   & \textbf{Description}\\
        \midrule
        $n_g$ &   Number of gateways in the system\\
        
        $n_i$ &   Number of IoT devices that connect to a gateway\\
        
        $n_r$   &   Number of random numbers an IoT device generates in one round of protocol\\
        
        $n_{mg}$ &   Number of malicious gateways in the system, and $m < n$\\
        
        $n_{mi}$ &   Number of malicious IoT devices connected to the same gateway, and $t < k$\\
        
        $\ell$  &   Number of source random numbers selected by the dApp for final random number generation\\
        \bottomrule
    \end{tabular}
\end{table}

% This part needs rewriting
\para{Problem statement}
\sln{} aims at guaranteeing generated numbers are uniformly distributed and unpredictable by an adversary under the security assumptions.
Specifically, \sln{} needs to address the following risks:
\begin{inparaenum}[\bf (i)]
    \item Compromised gateways. A gateway controlled by an adversary can abandon an unfavored random number generated by a connected IoT device, and refuse to respond according to the protocol based on the preference of the adversary;
    \item Compromised IoT devices. An IoT device (and the associated TEE) controlled by an adversary can generate biased random numbers, or discard a generated random number if the adversary cannot take advantage of it.
    %We assume an IoT device only interacts with a connected gateway and a compromised IoT device can create biased random numbers, or stop to create random numbers.
\end{inparaenum}

%\sln{} needs to guarantee that no one can manipulate the random number generation to gain advantages from the dApp execution.
%\item Malicious gateways. A dishonest gateway can discard one or more random numbers created by connected IoT devices, or completely stop response. A group of malicious gateways may also collude to influence the random number used by the dApp.

% Detailed protocol

\section{Detailed Random Number Generation Protocol}\label{sec-detailed-protocol}
We present the detailed design of \sln{} in this section.

\subsection{Random Number Generation on IoT Devices}
%To prevent a TEE device from dropping a request (intentionally or unintentionally), a generated random number is distributed 
we assume each of the IoT devices $d_i$ creates a random number $n_i$.
To facilitate the verification of the authenticity of $n_i$, the device $d_i$ also calculates a digital signature of the random number $\sigma_{d_i}(n_i)$.
We assume that the IoT device is equipped with TEE,  both the random number generation and signature generation are done inside the TEE and it is not likely that an attacker can manipulate these operations.

As the IoT device $d_i$ does not connect to the blockchain network directly, the generated message $(n_i, \sigma_{d_i}(n_i))$ is sent to the corresponding gateway for further processing.

\subsection{Random Number Processing on Gateways}
After receiving a message of a random number from a connected IoT device, the gateway cannot put it into the random number pool directly as the dApp may take advantage of it.
In other words, the numbers should only be visible to the dApp when it cannot make any choices.
To achieve this goal, \sln{} utilizes a commitment scheme, which satisfies two features:
\begin{inparaenum}[\bf (i)]
        \item Hiding. Hiding prevents one from learning information about the original message by observing the commitment.
        \item Binding. Binding guarantees that it is hard/impossible to open a commitment to a different value.
\end{inparaenum}
These two security features can be formalized in different flavors, such as computation hiding/binding and perfect hiding/binding.
The selection of a commitment scheme will not affect \sln{}, as long as the commitment scheme's security features are acceptable.
The only requirement is that the commitment scheme is non-interactive~\cite{naor1989bit}.
Specifically, the gateway utilizes a \textit{commitment scheme} to hide the received random number before sending it to the pool.
Each gateway runs the \texttt{Setup} when it joins the system. 
During this stage, the gateway runs the \texttt{Commit} algorithm on the generated random numbers one by one and submits all committed values to the pool.
The \textit{hiding} feature of the commitment scheme guarantees that no one can extract information from the original random number.
At the same time, the \textit{binding} feature of the commitment scheme prevents the gateway from modifying the committed value at a later stage.

The committed value has to be opened in a later stage when the dApp uses this particular random number.
The gateway prepares for the opening operation with a threshold secret-sharing scheme.
A secure $(k,t)$-threshold secret sharing scheme guarantees that with $t$ or more splits of the message, one can reconstruct the message.
But with less than $t$ splits, it is hard/impossible to rebuild the original message.
%Similar to the commitment scheme, a gateway also initializes the threshold secret sharing scheme with the corresponding \texttt{Setup} algorithm when it joins the system.
For each committed random number, the corresponding gateway uses the $(k, t)$-threshold secret sharing scheme to break the committed value into $k$ pieces, which are distributed to gateways in the system.
The values $k$ and $t$ are related to the security assumptions and system configuration, which will be discussed in \sectionname~\ref{sec-security-perf-analysis}.

\subsection{Random Number Request by dApp}
The dApp is responsible for generating a request for a random number, which is aggregated using multiple inputs from the pool for security purposes. 
The request must include two major parts that define the aggregation process, i.e., the way to select numbers in the pool and the method to aggregate selected numbers.
This request is distributed to all gateways in the system.
For security purposes, \sln{} verifies the request based on two criteria:
\begin{inparaenum}[\bf (i)]
    \item The request involves enough number of gateways. The minimum number of gateways needed in the request is determined by the security assumption, i.e., the number of gateways that are compromised and collude with the dApp in the system. 
    \item The aggregation algorithm only needs a few unbiased inputs to guarantee the output is unbiased. This criterion essentially means the aggregation algorithm needs to be able to tolerate biased inputs as much as possible. There are different ways to achieve this goal. The simplest one is treating each input as a binary string and using \texttt{XOR} operator to connect all operands, which only requires one unbiased operand to guarantee the output is unbiased. In most cases, a dApp does not have special requirements on the final random number except unbiasedness, and this simple \texttt{XOR}-based aggregation algorithm is enough.
\end{inparaenum}

\subsection{Response to the Request on Gateways}
After a request is created and submitted by the dApp, the gateways collaborate to help the dApp to compute the final random number.
The final random number is calculated through two steps:
\begin{itemize}
    \item Revealing committed random numbers in the pool. For a committed value, the opening information is shared with multiple gateways, and a subset of gateways can collaborate to open the committed random numbers. Because of the threshold feature, committed numbers can always be opened even if there is a small number of malicious gateways.
    \item Aggregating numbers in the pool to produce the final random number. This operation is straightforward when all involved elements in the pool are correctly opened. When the pool is accessible by the public, every blockchain node can easily verify the correctness of the final random number as the aggregation method is usually simple.
\end{itemize}

\subsection{Some Implementation Considerations}
We discuss the implementation of some components of \sln{}.
These design and implementation choices do not affect the architecture of \sln{}, though they may affect its performance and reliability.
%Several design decisions need to be made to put \sln{} into practice.

\para{Construction of the random number pool}
The random number pool serves as a temporary information storage system to hold generated random numbers of the current round (in the form of commitments).
It can also be used to store splits of commitments open information (each split should be cryptographically protected and only the corresponding).
There are several options to build this random number pool.
For instance, a centralized FTP server can work as a pool.
It is also possible to let each gateway to store everything, or utilize a decentralized storage system such as IPFS or other peer-to-peer storage mechanism~\cite{daniel2022ipfs} as the random number pool.
Each option for a random number pool has its pros and cons, and can be chosen based on the actual environment.

\para{The blockchain system}
\sln{} can work with all types of blockchain systems and dApps deployed atop.
As the blockchain is involved in the random number generation process (i.e., verification/opening of commitments and aggregation), its security features affect the reliability of these operations.
For instance, if the blockchain is constructed using a BFT style consensus protocol~\cite{long2019scalable}, an adversary controls more than 1/3 nodes participating the consensus can disable \sln{} or manipulate the final random numbers arbitrarily.

% Security and performance analysis

\section{Security and Performance Analysis}\label{sec-security-perf-analysis}
In this section, we analyze the generic construction of \sln{} from security and performance perspectives.

\subsection{Security Analysis}
There are two major strategies for the dApp and its colluders to manipulate the random number generation to maximize its benefits:
\begin{itemize}
    \item Active attack. For this type of attack, the dApp and its colluders actively modify the random number generation process to obtain preferred random numbers. Major active attacks include:
    \begin{inparaenum}[(i)]
        \item An IoT device always generates random numbers that the dApp prefers; and
        \item The dApp always selects gateways that collude with it in the request for a random number. 
    \end{inparaenum}
    
    \item Passive attack. For this type of attack, the dApp and its colluders do not actively change any information but skip certain numbers or ignore unfavorable requests. Major passive attacks include:
    \begin{inparaenum}[(i)]
        \item An IoT device discards a generated random number that is unfavorable for the dApp. While this attack requires the IoT device to actively skip generated numbers, the device does not need to modify anything. Therefore, we still classify this attack as a passive attack;
        \item A gateway refuses to collaborate to reveal committed numbers that are unfavorable for the dApp to force the protocol to terminate.
    \end{inparaenum}
\end{itemize}
\sln{} can prevent both types of attacks under the security assumptions discussed earlier.

\para{Prevention of Active Attack}
It does not matter what strategy the adversary takes, its ultimate goal is to manipulate one or more steps in the random number generation process to make the final value biased in a preferred manner.
\sln{} can thwart this type of attacks because it enforces a checking on the dApp's random number request, which guarantees two features:
\begin{inparaenum}[\bf (i)]
    \item The adversary cannot control all inputs. According to the security assumptions, only a subset of IoT devices/gateways can be compromised. \sln{} only needs to ensure that the aggregation algorithm involves more inputs than the maximum number values the adversary can control.
    \item The aggregation output cannot be influenced by a subset of inputs. Given an arbitrary aggregation algorithm, it is a hard problem to determine whether any single input can guarantee unbiasedness. But considering the normal requirements of the dApp on the final random number, \sln{} only needs to support \texttt{XOR} operation for aggregation, and this feature is preserved.
\end{inparaenum}

\para{Prevention of Passive Attacks}
The nature of a passive attack is ``denial-of-service'', i.e., when the adversary expects that an operation will lead to non-preferred results, it controls compromised entities to stop responding.
This type of attack is usually hard to prevent as it is not easy to force an entity to do certain things.
Most existing decentralized systems use an incentive mechanism to discourage such behaviors (e.g., offering rewards for taking actions and enforcing penalties for not taking actions). 
\sln{} adopts a different strategy to deal with this type of attack.
\begin{inparaenum}[\bf (i)]
    \item Prevention of passive attacks in the process of random number generation on the IoT device. From a high level, \sln{} divides the random generation into two stages. In the first stage, IoT devices/gateways contribute random numbers to a pool and the adversary does not know which will be included in aggregation. Therefore, the adversary does not have the motivation to control compromised IoT devices/gateways to refuse to participate in the protocol. 
    \item Prevention of passive attacks in the process of final random number generation. \sln{} prevents passive attacks in the aggregation stage by utilizing the threshold secret-sharing scheme to distribute the response capability to multiple entities. Based on security assumptions, there is always a subset of honest gateways. As long as the number of honest gateways exceeds the threshold, random numbers in the pool can be recovered.
\end{inparaenum}

\subsection{Performance Analysis}\label{subsec-perf-analysis}
In this section, we consider the performance of \sln{} with a generic commitment scheme and threshold secret sharing scheme.
A dApp has two performance requirements for \sln{}:
\begin{inparaenum}[(i)]
    \item Low latency. The latency is the period between the submission of the random number request and the receiving of the result. The dApp expects a quick response (i.e., low latency), especially when the dApp is time-sensitive. 
    \item Low cost. The on-chain computation is usually expensive and the cost is proportional to the amount of computation. To reduce the execution cost, the dApp expects to minimize the on-chain computation involved in the random number generation protocol.
\end{inparaenum}

\para{Computation cost analysis}
The generation of the final random number includes both on-chain and off-chain works.
While on-chain computation cost is more sensitive, it is better to minimize both on-chain and off-chain computation.

\subpara{Off-chain computation}
\sln{} keeps most of the computation off-chain, which includes three components:
\begin{inparaenum}[(i)]
    \item Random number commitment (\texttt{Commit}). Gateways convert all random numbers collected from IoT devices to the form of commitments. The total number of commit operations is $n_g \times n_i \times n_r$.
    \item Secret sharing (\texttt{Split}). Each gateway splits the opening information of each commitment and distributes to other $n_g - 1$ gateways. The total number of secret sharing operations is $n_g \times n_i \times n_r$.
    \item Secret recovering (\texttt{Reconstruct}). Only $\ell$ committed random numbers need to be recovered, so it involves $\ell$ \texttt{Reconstruct} operations.
\end{inparaenum}

\subpara{On-chain computation}
The on-chain part is mainly related to the aggregation phase, i.e., each blockchain node needs  which is further divided into two parts:
\begin{inparaenum}[(i)]
    \item Verification of numbers in the pool. This operation is equivalent to the opening of a set of committed values and the cost is determined by the commitment scheme, i.e., $\ell$ \texttt{Open} operations of the selected commitment scheme.
    \item Verification of aggregation process. The simplest way to verify the aggregation result is by repeating the process. If \texttt{XOR} is the only operator used for aggregation, the computation cost is $\ell-1$ \texttt{XOR} operations.
\end{inparaenum}

% under construction 
\para{Communication cost analysis}
There are two communication expensive steps in \sln{}, and both are related to the $(k,t)$-threshold secret sharing.
\begin{inparaenum}[(i)]
    \item Distribution of shares of committed random numbers. Each committed number is split into $k$ pieces, so the overall communication complexity is $\mathcal{O}(n_g \times n_i \times n_r \times k)$.
    \item Reconstruction of shares of committed random numbers. As only $\ell$ values need to be reconstructed, the overall communication complexity is $\mathcal{O}(\ell \times t \times k)$. Here we consider a worse situation, where each gateway distributes its splits to all $k$ other gateways. If all gateways are collaborative, this complexity can be reduced to $\mathcal{O}(\ell \times t)$.
\end{inparaenum}
%The first one is the distribution of shares of committed random numbers, and the complexity is $\mathcal{O}()$

% Improved protocol

\section{Improvement of \sln{}}\label{sec-improved-protocol}
The construction described in \sectionname~\ref{sec-detailed-protocol} is generic and does not consider the special features of the building blocks.

\para{Utilization of commitment scheme features}
While any secure commitment scheme can meet the security requirement of \sln{}, they are not equivalent from a performance perspective.
As analyzed in \sectionname~\ref{sec-security-perf-analysis}, the on-chain computation cost of the generic construction includes $\ell$ commitment open operation, and $\ell-1$ \texttt{XOR} operations which are employed to produce the final random number.
If the Pedersen commitment scheme~\cite{pedersen1991non} is adopted, each commitment open operation involves two exponents and one multiplication on the given finite field.
While this cost is not a big concern for a modern computer, it is not cheap as on-chain computation.

To further reduce the on-chain computation complexity, we utilize the additive homomorphic feature of the Pedersen commitment scheme.
Specifically, given two commitments $c_1 = g^{m_1}h^{r_1}$ and $c_2 = g^{m_2}h^{r_2}$, we can build the commitment of $m_1 + m_2$ with one multiplication, i.e., 
$$c_{1+2} = c_1 \cdot c_2 = g^{m_1 + m_2}h^{r_1 + r_2}.$$
With this feature, we can tweak the design of \sln{}, and a blockchain node does not need to open each committed value to verify its correctness before aggregating them to obtain the final random number.
Specifically, a blockchain node can aggregate the commitments utilizing the homomorphic feature first, and only verify the validity once.
With this strategy, the total on-chain computation cost is reduced to one aggregation of $\ell$ commitments ($\ell-1$ multiplications) and one open operation (two exponent operations and one multiplication).
Depending on the selection of $\mathbb{G}$, the exponent and multiplication can mean different arithmetic operations.
If $\mathbb{G}$ is a multiplicative group of a finite field, multiplication is a multiplication of two big integers modulo a prime number.
If $\mathbb{G}$ is a group of elliptic curve points, multiplication is a scalar multiplication of elliptic curve points.
In both cases, the complexity of a typical efficient exponent operation is equivalent to $\mathcal{O}(\log |\mathbb{G}|)$ multiplication operations.
In other words, the exponent operation is much more expensive than multiplication, and the new approach can greatly reduce the on-chain computation complexity.
We analyze the security and performance of the improved \sln{} in the rest of this section.

\para{Security analysis of improved \sln{}}
With the improved \sln{},  the blockchain can only obtain the addition of $\ell$ random numbers
\begin{equation}\label{equ-add-result}
    m_1 + \cdots + m_{\ell},
\end{equation}
not the original bitwise \texttt{XOR} result
\begin{equation}\label{equ-xor-result}
    m_1 \oplus \cdots \oplus m_\ell.
\end{equation}

Without loss of generality, we assume each $m_i$ has the same bit length.
For \equref{equ-xor-result}, as long as one of the input values $m_i, 1 \leq i \leq \ell$ is an unbiased random bit string, the aggregation result is an unbiased random value.
The situation of \equref{equ-add-result} is more complex.
If all $m_i$s are defined as integers within a given range and selected randomly, the sum of these values follows a variant of Irwin-Hall distribution~\cite{hall1927distribution,irwin1930frequency} other than uniform distribution in that range.
Fortunately, $m_i$s are defined in a finite group that is isomorphic to $\mathbb{Z}/p\mathbb{Z}$ where $p$ is a prime number (actually we have $|\mathbb{G}|=p$). 
Accordingly, the addition given in \equref{equ-add-result} is addition modulo $p$.
Addition in this structure offers a similar feature as \texttt{XOR} on bit strings.
Specifically, as long as one element $m_i$ is uniformly selected from $\mathbb{Z}/p\mathbb{Z}$, the result of \equref{equ-add-result} is uniformly distributed in $\mathbb{Z}/p\mathbb{Z}$.

To prove this feature, we assume $m_1$ is randomly selected from $\mathbb{Z}/p\mathbb{Z}$, and the adversary manages to set the sum of other elements as a preferred value $a \in \mathbb{Z}/p\mathbb{Z}$. Given an arbitrary value $b \in \mathbb{Z}/p\mathbb{Z}$,
\begin{equation}\label{equ-uniform-proof}
    \Pr[m_1 + a = b] = \Pr[m_1 = b-a] = \frac{1}{p}.
\end{equation}
\equref{equ-uniform-proof} states that it does not matter what strategies the adversary uses to choose the value $a$, the addition result will be uniformly distributed among all the possible values in $\mathbb{Z}/p\mathbb{Z}$. 
%
%Therefore, the addition result is still uniformly distributed in $\mathbb{Z}/p\mathbb{Z}$.

Note that there is a tiny difference between a random number in $\mathbb{Z}/p\mathbb{Z}$ and a random bit string in $\{0,1\}^{\lceil \log (p) \rceil}$.
Hardware-based TRNGs usually create random bit strings other than random elements in $\mathbb{Z}/p\mathbb{Z}$.
To convert a binary string to an element of $\mathbb{Z}/p\mathbb{Z}$, we can simply treat the bit string as an integer and apply a modulo $p$ operation.
When the length of the bit string is longer than ${\lceil \log (p) \rceil}$, the converted result roughly follows uniform distribution in $\mathbb{Z}/p\mathbb{Z}$. 
Given a value $p$, a longer bit string can lead to a more uniform distribution in $\mathbb{Z}/p\mathbb{Z}$.
If the dApp needs a random binary string other than an element in $\mathbb{Z}/p\mathbb{Z}$, we can treat the element as a binary string and reduce the leading bits to guarantee it roughly follows the uniform distribution.

%Another potential risk is that the aggregated 

%From security perspective, these two aggregation approaches are not equivalent, and the addition of multiple uniformly selected random numbers follows a variant of Irwin-Hall distribution~\cite{hall1927distribution,irwin1930frequency} other than uniform distribution, which can be a potential risk for the safety/fairness of the dApp that relies on the random number.

%A variable that follows the Irwin-Hall distribution can be converted to a uniform distribution using inverse transform sampling~\cite{devroye1986sample}, but the computation process is too complex for on-chain operation.
%To close this gap, a cryptographic hash function is applied to the addition result and the hash value is used as the final random number.
%Under the random oracle model, a secure hash function is used to mimic a random oracle, i.e., given an input, it outputs a uniformly distributed bit string.
%Therefore, the use of the cryptographic hash function can efficiently eliminate the bias of the addition result.

%Note that this approach (i.e., applying a hash function to a biased input to produce a uniformly distributed random value) does not work without other components of \sln{} mainly because of denial-of-service attack, i.e., a malicious entity can compute the hash value by itself and refuse to participate the final random number generation if it finds out the hash output is not desirable.

\para{On-Chain performance analysis of the improved \sln{} on Ethereum}
The improved \sln{} relies on a smart contract to open/verify a set of Pedersen commitments and aggregate all the random numbers using modular additions over a prime finite field. 
Here we assume the Pedersen commitment scheme is built on an elliptic curve points group, which is common in blockchain.
\tablename~\ref{tbl-gas} summarizes the gas cost for performing arithmetic operations on the Ethereum Virtual Machine (EVM), where the gas cost for computing point addition and scalar multiplication is based on the precompiled contracts on the \texttt{alt\_bn128} elliptic curve~\cite{eip1108}.  

\begin{table}
    \centering
    \caption{Gas cost for performing arithmetic operations in EVM of Ethereum~\cite{evmcodes}}
    \label{tbl-gas}
    \begin{tabular}{p{2in}p{1in}p{1in}}
        \toprule
        \textbf{Arithmetic Operation} & \textbf{EVM Opcode} & \textbf{Gas Cost} \\
        \midrule
        Modular Addition & \texttt{addmod} & $8$\\
        Modular Multiplication & \texttt{mulmod} & $8$\\
        EC Point Addition & \texttt{ecadd} & $150$\\
        EC Scalar Multiplication & \texttt{ecmul} & $6,000$\\
        \bottomrule
    \end{tabular}
\end{table}

\tablename~\ref{tbl-evaluation} provides a comparison of the gas cost between the non-optimized \sln{} construction and the optimized one as described in \sectionname~\ref{sec-detailed-protocol} and \ref{sec-improved-protocol}, respectively. Thanks to the homomorphic property of the Pedersen commitment scheme, it is not difficult to find that the optimized implementation of \sln{} can save the gas cost significantly with the increasing of aggregated random numbers.
Note that the gas cost is directly associated with computation complexity, i.e., a method is more expensive in gas if it requires more arithmetic operations.

\begin{table}
    \centering
    \caption{Performance evaluation and gas cost for opening and aggregating $\ell$ random numbers in EVM of Ethereum.}
    \label{tbl-evaluation}
    
    \begin{tabular}{p{1in}p{0.65in}p{0.65in}p{0.65in}p{0.75in}}
        \toprule
        \multirow{2}{*}{\textbf{\sln{}}} & \multicolumn{3}{c}{\textbf{Operation Calls}} & \multirow{2}{*}{\textbf{Gas Cost}} \\
        
               &\#\texttt{ecadd} & \#\texttt{ecmul} & \#\texttt{addmod} & \\
        \midrule
        Non-optimized & $\ell$ & $2\cdot\ell$ & $\ell - 1$  & $12,158\cdot\ell - 8$\\ 
        
        Optimized     & $\ell$ & $2$ & $2\cdot(\ell - 1)$  & $166\cdot\ell + 11,984$\\ 
        \bottomrule
    \end{tabular}

\end{table}

\begin{figure}
  \centering
  \includegraphics[width=0.4\linewidth]{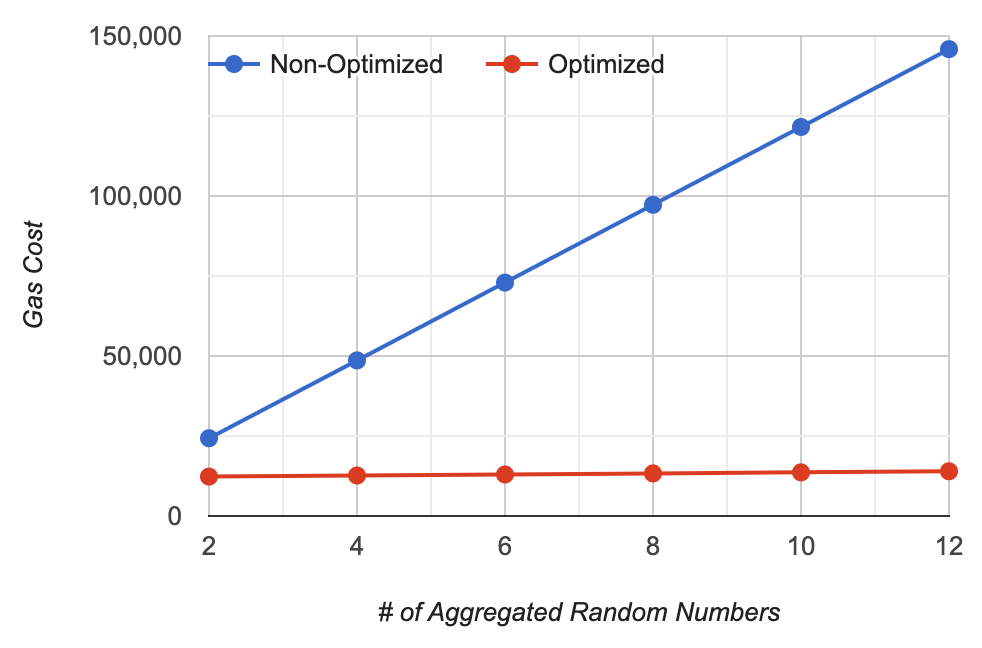}
  \caption{The gas costs for non-optimized and improved \sln{}. Here both versions use the elliptic curve-based Pedersen commitment scheme but the unoptimized \sln{} does not utilize the homomorphic feature. The $x$-axis is the number of values to be aggregated and the $y$-axis is the amount of gas required to finish the work (commitment openings and aggregation).}
  \label{fig:gas_cost}
\end{figure}

\figurename~\ref{fig:gas_cost} shows the gas costs of the two versions of \sln{}.
It is easy to see that the gas cost of the non-optimized \sln{} grows linearly as more random numbers are used to produce the final result, while the gas cost of the improved \sln{} is almost constant.
This difference is important from security perspective, as it is harder for an adversary to manipulate the final result when more numbers are involved.
When the system uses 12 random numbers to produce the final result, the improved \sln{} only consumes about $10\%$ of the gas of \sln{} without optimization.

%24308, 48624, 72940, 97256, 121572, 145888
%12316, 12648, 12980, 13312, 13644, 13976

%l*150 + 2l*6000 + 8*(l-1)  
%l*150 + 2*6000 + 16*(l - 1)

%\para{Security analysis of improved \sln{}}
%The improved \sln{} only changes the way to verify the integrity of the source random numbers and the 

%\para{Performance evaluation of improved \sln{}}

% Conclusion

\section{Conclusion}\label{sec-conclusion}
Random numbers play a critical role in blockchain and dApps, and various random number generation mechanisms have been designed to meet this demand.
Compared with previous methods, \sln{} works in a hybrid manner, i.e., it utilizes both hardware-based TRNG and on-chain/off-chain algorithms for random number generation.
This hybrid approach allows \sln{} to enjoy the benefits of different random number generation strategies and can tolerate various malicious participants.
As on-chain computation is usually expensive, we also design an efficient on-chain aggregation mechanism that does not sacrifice the quality of generated random numbers and decentralization level.
The analysis and evaluation show that this improvement can save significant cost (both computation burden and gas fee if deployed for Ethereum) compared with the non-optimized construction.

% Bibliography
\balance
\bibliographystyle{plainnat}
\bibliography{ref}

\end{document}